\documentclass[trackchanges,twocolumn,resetfootnote]{aastex701}

\usepackage{amsmath}

\revised{\today}
\usepackage{appendix}
\usepackage{subfigure}
\usepackage{stfloats}
\usepackage{afterpage}

\graphicspath{{./}{figure/}}

\begin{document}
\title{Quasi-Periodic Polarized Emissions from Kink Structure in Magnetized Relativistic Jets}

\author[orcid=0009-0006-9884-6128]{Xu-Fan Hu}
\email{ebr105@sjtu.edu.cn}
\affiliation{Tsung-Dao Lee Institute, Shanghai Jiao-Tong University, Shanghai, 1 Lisuo Road, 201210, People's Republic of China}

\author[orcid=0000-0003-0292-2773]{Hong-Xuan Jiang}
\email{hongxuan\_jiang@sjtu.edu.cn}
\affiliation{Tsung-Dao Lee Institute, Shanghai Jiao-Tong University, Shanghai, 1 Lisuo Road, 201210, People's Republic of China}

\author[orcid=0000-0002-8131-6730]{Yosuke Mizuno}
\email{mizuno@sjtu.edu.cn}
\affiliation{Tsung-Dao Lee Institute, Shanghai Jiao-Tong University, Shanghai, 1 Lisuo Road, 201210, People's Republic of China}
\affiliation{School of Physics \& Astronomy, Shanghai Jiao-Tong University, Shanghai, 800 Dongchuan Road, 200240, People's Republic of China}
\affiliation{Key Laboratory for Particle Physics, Astrophysics and Cosmology (MOE), Shanghai Key Laboratory for Particle Physics and Cosmology, Shanghai Jiao-Tong University, 800 Dongchuan Road, Shanghai, 200240, China}

\author[orcid=0000-0002-1827-1656]{Christian M. Fromm}
\email{christian.fromm@uni-wuerzburg.de}
\affiliation{Institut f\"ur Theoretische Physik und Astrophysik, Universit\"at W\"urzburg, Emil-Fischer-Str. 31, D-97074 W\"urzburg, Germany}
\affiliation{Institut f\"ur Theoretische Physik, Goethe-Universit\"at Frankfurt, Max-von-Laue-Stra{\ss}e 1, D-60438 Frankfurt am Main, Germany}
\affiliation{Max-Planck-Institut f\"ur Radioastronomie, Auf dem H\"ugel 69, D-53121 Bonn, Germany}

\author[orcid=0000-0001-5424-0059]{Bhargav Vaidya}
\email{bvaidya@iiti.ac.in}
\affiliation{Discipline of Astronomy, Astrophysics and Space Engineering, Indian Institute of Technology Indore, Khandwa Road, Simrol, 453552, India}

\correspondingauthor{Xu-Fan Hu, Yosuke Mizuno}
\email{ebr105@sjtu.edu.cn, mizuno@sjtu.edu.cn}

\begin{abstract}
Recent polarimetric observations of blazars indicate the development of current-driven (CD) kink instability after passing the recollimation shocks in the relativistic jets and association with quasi-periodic oscillations (QPOs). 
To investigate multi-wavelength polarized features of CD kink instability in jets, we develop {\tt RaptorP}, a new special relativistic module of the polarized General Relativistic Radiative Transfer (GRRT) code {\tt RAPTOR}.
Based on 3D SRMHD simulations of over-pressured magnetized jets, we find that jet images vary at different frequencies. At low frequencies, the emission comes from the turbulent ambient medium surrounding the jet that obscures the inner jet structure. Electronic Vector Position Angle (EVPA) patterns are perpendicular to the jet propagation direction, indicating a dominance of the poloidal magnetic field.
At high frequencies, bright knots and twisted kink structures appear, and EVPA patterns are consistent with a toroidal magnetic field. 
We also find that QPOs in light curves of intensity and linear polarization (degree and angle). The peak frequency in Power Spectral Densities (PSDs) is well-matched with the rotation period of the kink structure in relativistic jets. It shows an anti-correlation between total intensity and the degree of polarization at a lower inclination angle.
Our findings, based on realistic polarized radiation calculations, will explain the observational signatures seen in blazars.

\end{abstract}


\section{Introduction} \label{intro}

Relativistic jets are widespread in Microquasars, Active Galactic Nuclei (AGNs), and Gamma-Ray Bursts (GRBs). 
Great efforts have been made on searching jets from radio to X-ray band, with more than reported $\sim10^9$ extragalactic radio sources, in low-frequency radio surveys \citep[e.g.,][]{blandford2019}.
They maintain highly collimated over vast distances, spanning from sub-parsec to mega-parsec scales \citep[e.g.,][]{pushkarev2009}, with recent observations showing they can even reach the cosmic web \citep{oei2024}.

Blazars are AGNs with highly relativistic jets nearly along our line of sight, which exhibit highly variable, non-thermal emission from radio to TeV $\gamma$-rays.
Their emission is strongly polarized, with a typical degree of polarization $\sim0.1$-$0.2$ \citep{kang2015,jortstad2022,liodakis2022,sciaccaluga2025}. 
While shock acceleration has long been considered the dominant particle acceleration mechanism in blazar jets \citep{bottcher2010}, recent studies suggest that such models struggle to account for the observed rapid variability and polarization properties \citep{sironi2015,zhang2017}. 
Instead, magnetic reconnection and turbulence driven by current-driven (CD) kink instabilities emerge as plausible alternatives, providing a natural explanation for the power-law distributions of non-thermal particles \citep{alves2018,davelaar2020}. 
Qusai-Periodic Oscillation (QPO) is widely observed from radio to $\gamma$-ray band in blazars \citep{tripathi2021,jortstad2022,sharma2025}, characterized by the peak in the Power Spectral density (PSD). \cite{jortstad2022} found QPO at optical flux and linear polarization, and $\gamma$-ray flux in BL Lacarte, with a period around 13~hours. The degree of polarization at {\tt R}-band oscillated between 0 and 0.15. The observation image matches the hypothesis that the CD kink instability happens after passing the recollimation shocks.

Special relativistic magnetohydrodynamic (SRMHD) simulations have been conducted to understand complex structures developed by shocks and instabilities in relativistic jets.
Simulations of over-pressured jets in two-dimensional SRMHD simulations show the development of recollimation shocks, which explain the quasi-stationary emission features seen in observations \citep{mizuno2015,thimmappa2024}. 
Besides, various instabilities in jets have been found and studied, such as CD kink instability \citep{mizuno2009,mizuno2014}, Kelvin-Helmholtz instability (KHI) \citep{millas2017,matsumoto2019}, Centrifugal instability (CFI) \citep{gourgouliatos2018}, Rayleigh-Taylor instability (RTI) \citep{toma2017,abolmasov2023}, etc. Particularly, the CD kink instability arises when a jet with a strong toroidal magnetic field experiences a non-axisymmetric perturbation, resulting in a characteristic twisted structure. Recently, \cite{hu2025} extended \cite{mizuno2015} into 3D SRMHD simulations, where the development of CD kink instability in recollimation shocks has been identified in relativistic jets, providing a potential explanation for the moving kink structure seen in BL Lac\citep{jortstad2022}.

To investigate how the development of kink instability drives QPOs, introducing radiation transfer into the simulation process is necessary. A common approach is to post-process simulated data by assuming an electron distribution function (eDF), such as a power-law with exponential cutoff, to calculate the local emission in each cell and then integrate total intensity along line of sight \citep{dong2020,dubey2023,dubey2024}.
For the first time, \cite{dong2020} showed that the kink instability could drive QPO. 
\cite{upreti2024} utilized Lagrangian particles \citep{vaidya2018} embedded in {\tt PLUTO} to evolve a power-law eDF and also calculated radiation through this approach. They applied the kink instability to explain the rib-like structure in radio galaxy {\tt MysTail}.
This approach omitted the effects of optical depth, making the results valid only at high frequencies where the plasma is optically thin. At low frequencies, this method is inaccurate for blazars because synchrotron self-absorption can be important for their smaller emission regions and stronger magnetic fields. 
Meanwhile, the absence of Faraday rotation calculation prevents direct comparison with electric-vector position angles (EVPAs) from real observations.

Beyond the limitation associated with radiative transfer effects, a further significant concern regarding the aforementioned methodologies are somehow dependence on the assumption of a power-law eDF\footnote{Lagrangian particle module evolves the particle spectra with a power-law initialisation accounting for radiative losses and energisation at shocks. }. Given that the emissivity of the power-law is independent to electron temperature and density, the integral emission is predominantly produced by the external ambient gas rather than the internal jet components. In \cite{kramer2024}, the ambient medium entirely obscured the jet. By applying a jet tracer to exclude the ambient gas, they revealed the jet emission. However, the ambient emission is underestimated.

In this work, we present the new module of Special Relativistic Radiative Transfer {\tt RaptorP}, based on the polarized General Relativistic Radiative Transfer (GRRT) code {\tt RAPTOR} \citep{bronzwaer2018,bronzwaer2020}. 
Non-thermal emission is incorporated via sub-grid models using prescriptions from recent PIC simulations. 
By employing the newly developed SRRT code {\tt RaptorP}, we extend \cite{hu2025} to generate multi-frequency images and spectral energy distribution (SED) of a simulated over-pressured magnetized jet. Through this, we identify distinct radiative polarized signatures across various frequencies and investigate QPOs in light curves.

The structure of this paper is as follows. In Section~\ref{rmhd}, we review the setup and results of the simulation of the over-pressured magnetized jet. In Section~\ref{srrt}, we describe the features of the new SRRT code. In Section~\ref{re}, we compare radiation images among different viewing angles, eDFs, and frequencies. In Section~\ref{tsa}, we analyze light curves from the kink region and seek QPOs. We summarize our findings in Section~\ref{con}. 

\section{Method} \label{meth}

\subsection{Setup of SRMHD simulation}\label{rmhd}

\begin{figure}
	\centering
	\includegraphics[width=0.47\columnwidth]{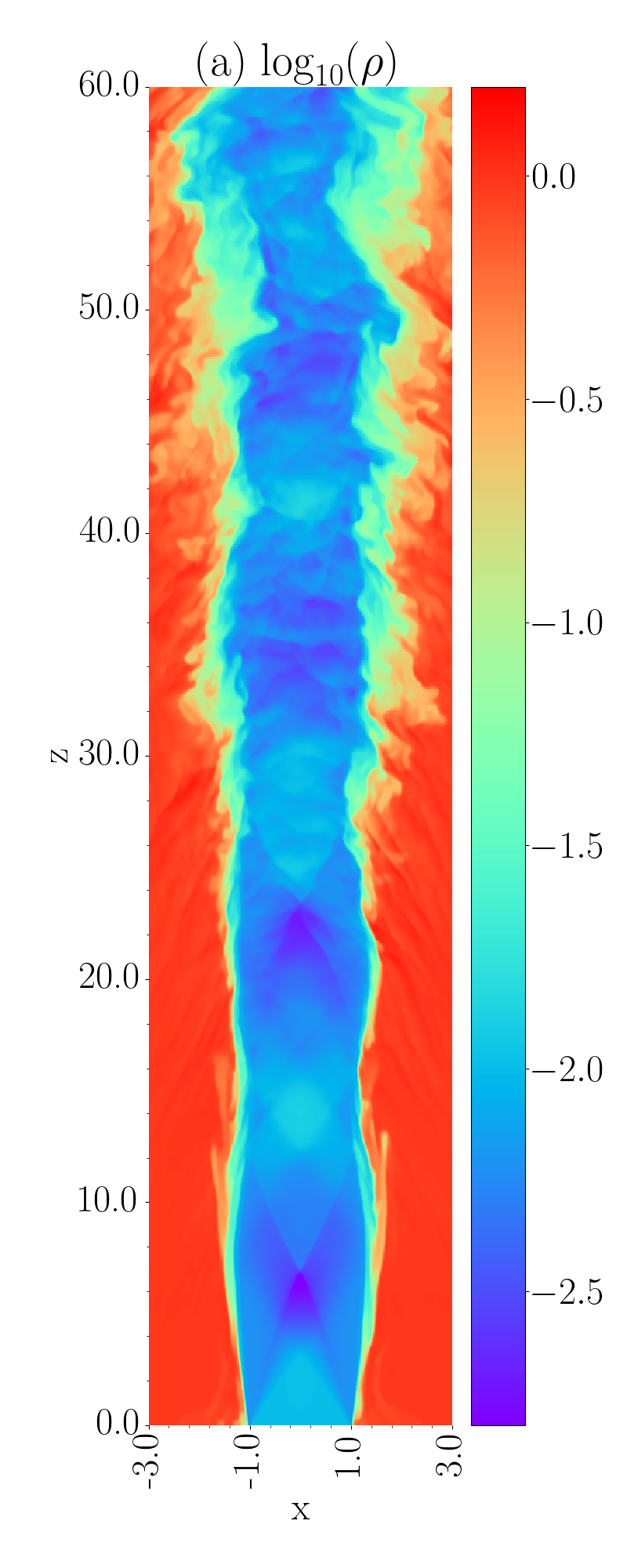}
	\includegraphics[width=0.47\columnwidth]{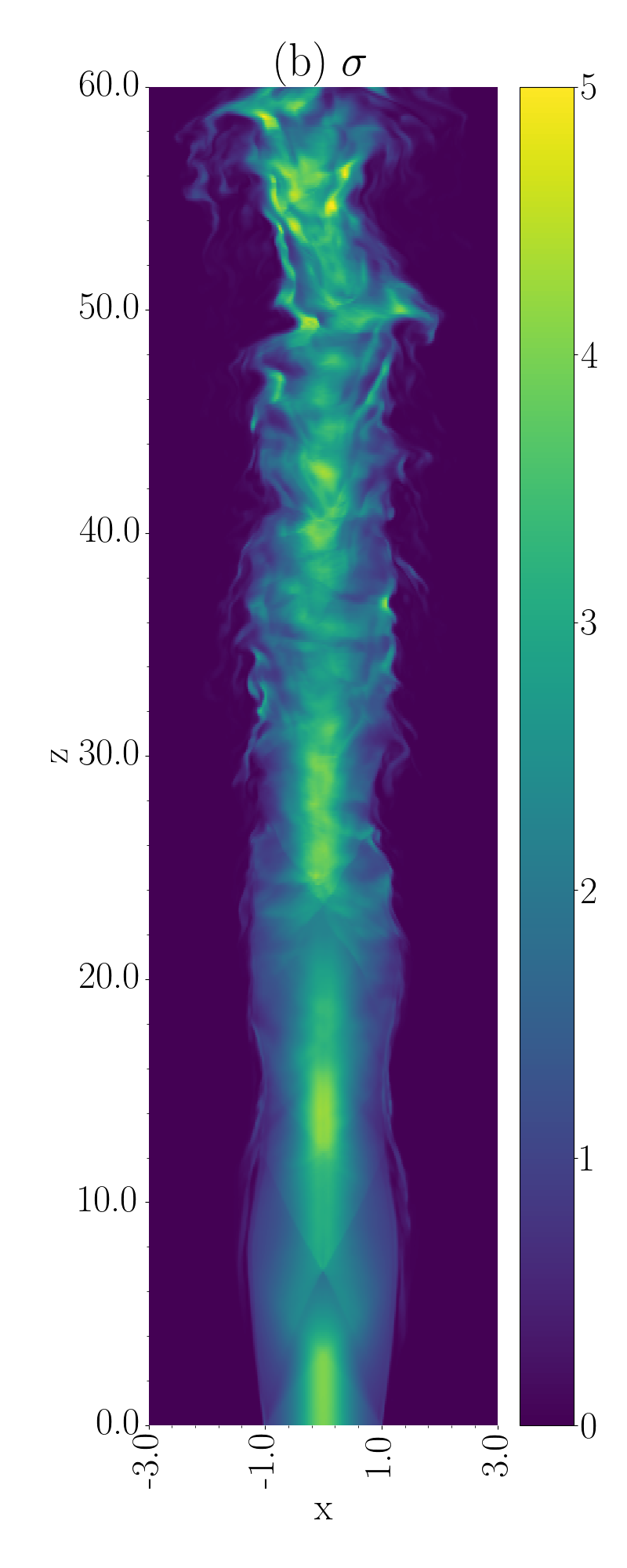}
	\caption{2D axial distribution of (a) logarithmic density, (b) magnetization of over-pressured magnetized jet at $t_s=400$.}
	\label{fig:rmhd}
\end{figure}

Following the setup of \cite{hu2025}, we simulate 3D over-pressured jets by solving the SRMHD equations with the {\tt PLUTO} code \citep{mignone2007}. 

We assume a preexisting cylindrical over-pressured jet. 
The jet radius $R_j$ is set to be unity. It has an axial velocity with Lorenz factor $\gamma_j=3$, which corresponds to $v_j=0.9428c$. To trigger the CD kink instability, a toroidal-dominated helical magnetic field is adopted. ({\tt MHD3} setup, see \cite{hu2025} for details). Recent GRMHD simulations of large-scale kink-unstable jets \citep[e.g.,][]{lalakos2024} achieve a magnetization $\sigma=b^2/\rho$ around $1-10$, where $b^2$ is square of 4-magnetic fields and $\rho$ is density. In this work, we set the magnetization in the range of the above simulations, which is between 2 and 4.  
The ambient medium affects the development of CD kink instabilities. Recent studies identify that a shallow ambient density profile makes the jet less stable for kink mode \citep[e.g.,][]{tchekhovskoy2016,duran2017,bodo2022}. In this study, to simplify the model, we adopt a uniform stationary ambient environment. 

The computational domain is $(x,y,z)=(\pm 5R_j,\pm5R_j, 60R_j)$ with a uniform grid of $(N_x,N_y,N_z)=(500,500,600)$. 
We impose outflow boundary conditions on all surfaces except for $z=0$. At $z=0$, we use fixed boundary conditions that continuously inject the over-pressured jet into the computational domain. 

In \cite{hu2025}, the development of CD kink instability has been systematically investigated. Here, we demonstrate the CD kink instability from the RMHD simulation, which is shown in Figure~\ref{fig:rmhd}. Density $\rho$ and magnetization $\sigma$ distributions at $t=400 t_s$ are presented, where the simulation time unit is $t_s=R_j/c$. At this time, the system has reached a quasi-stable state. At $z<30 R_j$, due to the mismatch of pressure between the jet and ambient medium, the recollimation shocks are developed and seen in panel (a). A CD kink instability grows in the jet at $z>40R_j$, showing a highly twisted structure.

\subsection{SRRT code}\label{srrt}

{\tt RAPTOR} \citep{bronzwaer2018,bronzwaer2020} solves the GRRT calculation through two main steps: determining the null geodesic and integrating the effect of radiative transfer (see appendix \ref{rt}). In this work, we assume jets are sufficiently far from black holes that general relativistic effects are ignored. 
Here, we newly develop {\tt RaporP}, the SRRT extension of the polarized GRRT code {\tt RAPTOR}. This implementation performs radiative transfer calculations in Minkowski spacetime under the fast-light approximation.

{\tt RaptorP} code can read both uniform and non-uniform Cartesian grid data of {\tt PLUTO}. It also supports a tracer to exclude the ambient medium. From one-sided jet simulation data, it can generate a symmetric two-sided jet. The source code and tutorial are publicly available on an open-access repository\footnote{\url{https://github.com/Lace-t/RaptorP}}.

In relativistic jets, non-thermal emission is essential \citep[e.g.,][]{Cruz-Osorio2022a,fromm2022,Yang2024}. 
To capture the microphysics of particle acceleration ignored in macroscopic MHD, we implement sub-grid prescriptions for non-thermal emission. We separately adopt fitting formulas from PIC simulations of turbulent plasma \citep{meringolo2023} and magnetic reconnection \citep{ball2018}, which provide estimations for both the power-law index and acceleration efficiency.

The efficiency in the production of non-thermal particles in terms of the weighted average of the excess over a Maxwell–J\"{u}ttner distribution is defined as:
\begin{equation}
    \epsilon=\frac{\int_{\gamma_{0}}^{\infty}(dN/d\gamma-f_{MJ}(\gamma,\Theta_e))(\gamma-1)d\gamma}{\int_{\gamma_{0}}^{\infty}(dN/d\gamma)(\gamma-1)d\gamma},\label{10}
\end{equation}
where 
\begin{equation}
f_{MJ}(\gamma,\Theta_e)=\frac{\gamma^2v}{c\Theta_eK_2(1/\Theta_e)}e^{-\gamma/\Theta_e}
\end{equation}
with $K_2$ the modified Bessel function of the second kind, and $\gamma_{0}$ denotes the peak of spectrum. We tie it with the peak of thermal distribution $\sim1+3\Theta_e$ \citep{chatterjee2021}.

As an example, the efficiency and power index estimated from turbulent plasma in \cite{meringolo2023} are

\begin{align}
    \epsilon &=1-\frac{0.23}{\sqrt{\sigma}}+0.5\sigma^{0.1}\tanh(-10.18\sigma^{0.1}\beta)\\
    \kappa &=2.8+\frac{0.2}{\sqrt{\sigma}}+1.6\sigma^{-0.6}\tanh(2.25\beta\sigma^{1/3})\label{12}
\end{align}

Then we model the non-thermal emission with $\kappa$ eDF \citep{xiao2006,davelaar2019}: 
\begin{equation}
	\frac{d N}{d\gamma} \propto \gamma\sqrt{\gamma^2-1}(1+\frac{\gamma-1}{\kappa w})^{-(\kappa+1)}. \label{13}
\end{equation}

To eliminate dependence on an arbitrary parameter \citep{davelaar2019},  we derive a self-consistent determination of $w$ based solely on $\epsilon$ and $\kappa$ (see Appendix \ref{w}).

\begin{figure*}
	\centering	
    \includegraphics[width=0.95\linewidth]{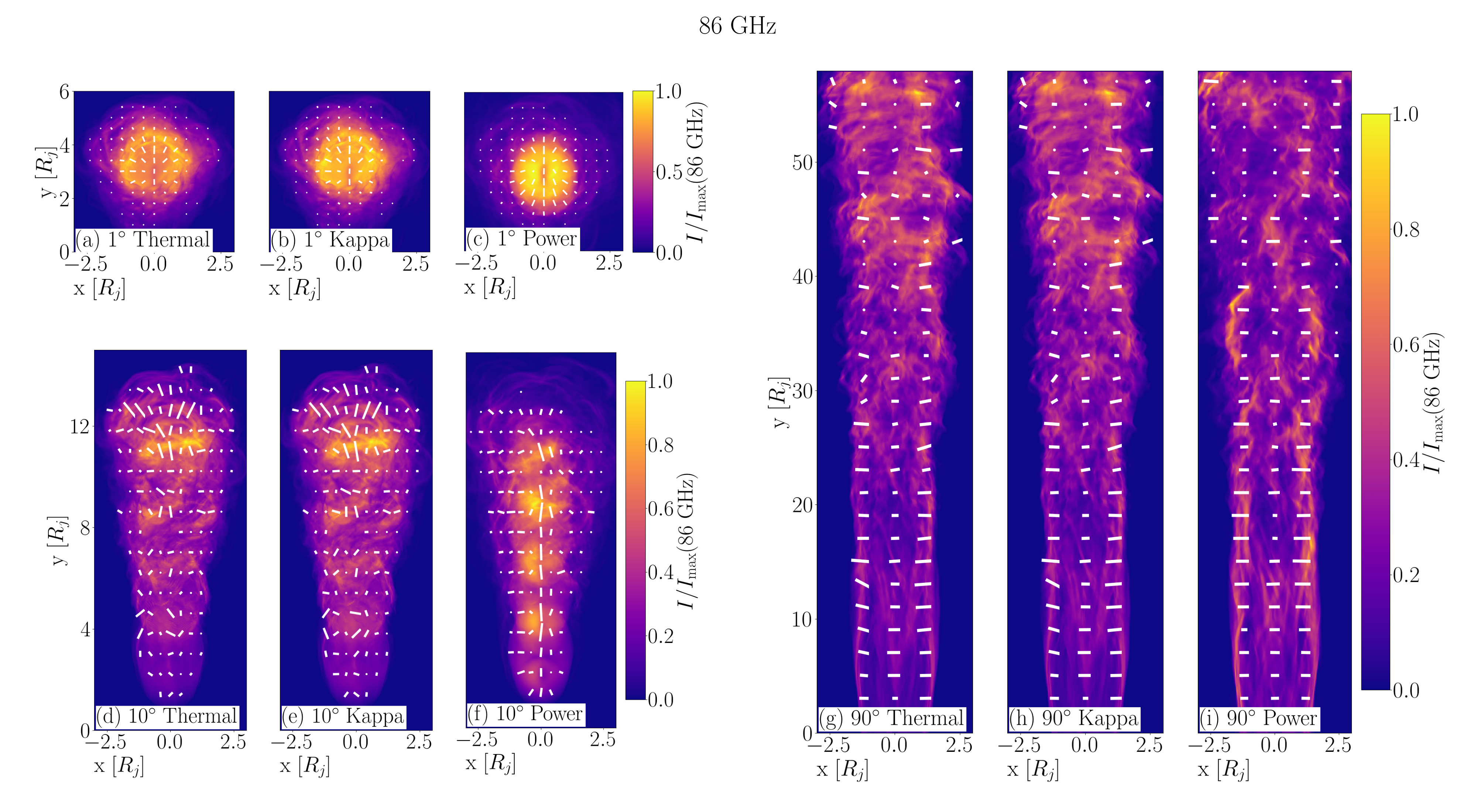}
	\caption{Normalized ray-Traced snapshot images of over-pressured magnetized jet of 86~GHz frequency at $t_s=400$ with inclination angles $i=1^\circ$ ({\it left upper}) , $10^\circ$ ({\it left lower}) and $90^\circ$ ({\it right}) for  thermal ({\it left}), $\kappa$ ({\it middle}), and broken power-law distribution ({\it right}). In each panel, we normalize with the maximum intensity. White lines mark EVPA patterns (shown only degrees of polarization exceeding 10\%).}
	\label{fig:angle}
\end{figure*}

\begin{figure*}
	\centering
	\includegraphics[width=0.8\linewidth]{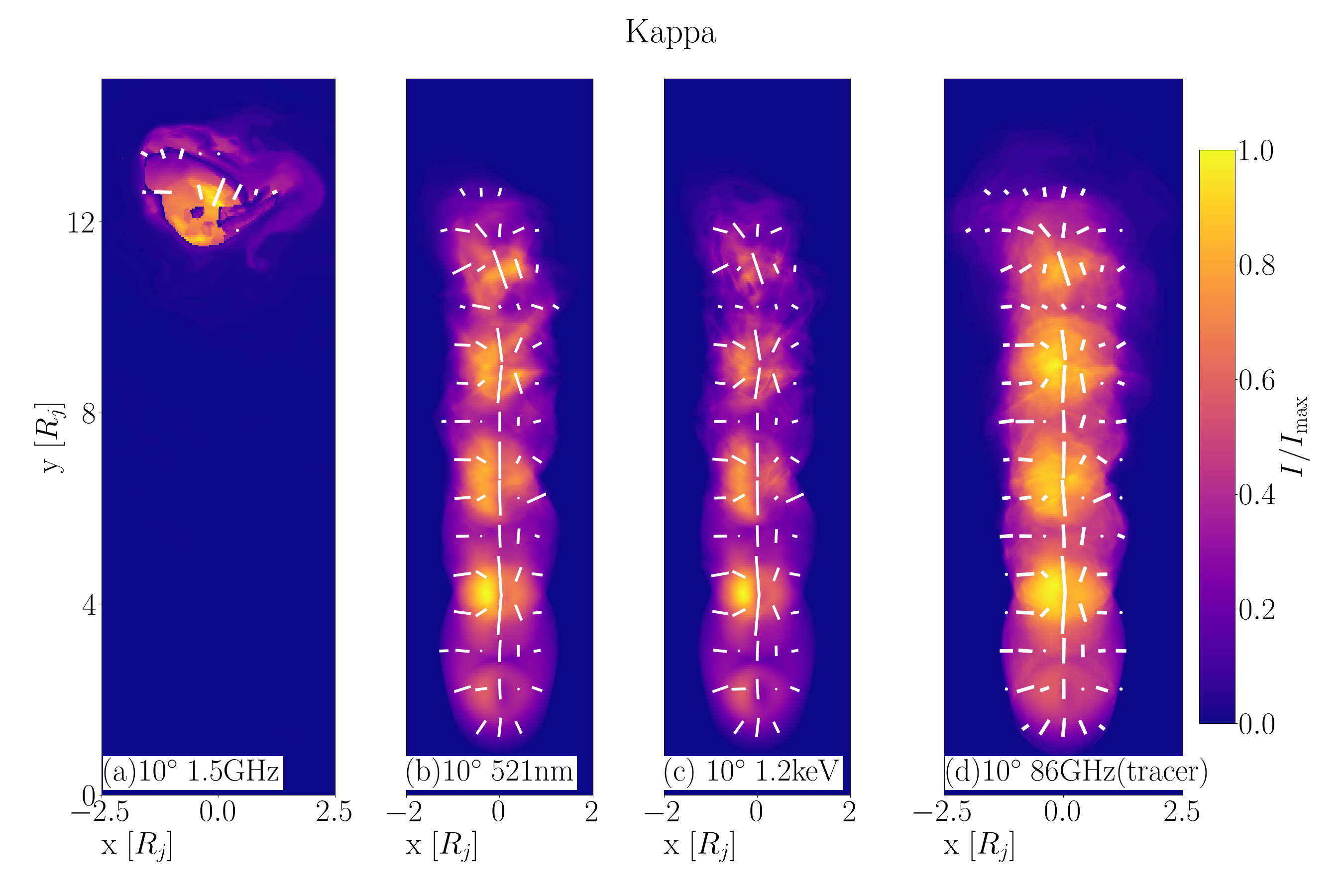}
	\caption{Ray-Traced intensity images of over-pressure magnetized jet at $t_s=400$ with an inclination angle $i=10^\circ$ at the frequencies of (a) 1.5~GHz, (b) 521~nm, (c) 1.2~keV, (d) 86~GHz with a tracer to exclude the environment. Here, we use the $\kappa$ distribution.}
	\label{fig:freq}
\end{figure*}

In this work, we compare three different eDFs: Maxwell-J\"uttner(thermal), $\kappa$(hybrid), and broken power-law(non-thermal) distributions. For the power-law distribution, we assume $\gamma_{min}=1$ and the power-law index is fixed at 2.3, following the PIC simulation results of \cite{alves2018}, since the magnetization parameter ($\sigma$) in the jet ranges between 2 and 4 (Figure~\ref{fig:rmhd} (b)). 

Lastly, to run {\tt RaptorP}, we choose arbitrary units for the black hole mass and plasma density. Here we set $M_{\rm BH}=6.5\times10^7 M_\odot, M_{\rm unit}=1\times10^{26} g$. The length unit is set as $L_{unit}=R_j=100GM_{BH}/c^2$. Applied these units, the magnetic field strength is $1$-$10$~Gauss in the jet, and the jet power is $3.1\times10^{42}\ \mathrm{erg/s}$.

\section{Result} \label{re}

\subsection{Influence of eDFs and viewing angles}

First, we investigate polarized radiation images of relativistic jets.
To mimic the observation of jets in Blazars as well as the other types of AGNs,
we simulate 86~GHz jet images of over-pressured magnetized jets at $t=400 t_s$ in three distinct viewing angles: $1^\circ$, $10^\circ$, and $90^\circ$ seen in Figure~\ref{fig:angle}.
The camera's field of view (FOV) and resolution are adjusted at different inclination angles to obtain fine images.
These calculations allow for a direct comparison of our results with previous work \citep{kramer2021, dubey2024}. In Figure~\ref{fig:angle}, the intensity of each panel is normalized with the maximum intensity. The EVPA patterns are plotted where degrees of polarization exceed 10\%. Thus, they mark the linearly polarized regions.

We find that qualitatively, all of three different eDFs exhibit similar distributions of intensity map and EVPA patterns. 
Ray-traced images at a viewing angle of $45^\circ$ are also calculated (see Appendix~\ref{45}), and are similar to $90^\circ$. 

At a viewing angle of $1^\circ$, the radiation signatures can be divided into two parts according to intensity and polarization. The inner jet shows strong emissions and radial EVPA patterns that indicate the existence of a toroidal magnetic field. The outer jet, which is a turbulent interface region, is dim and unpolarized.

At $10^\circ$ and $90^\circ$, the global profiles of the emission can also be divided into two parts: the upper kink region has the emission from the turbulent medium surrounding the jet, showing stronger intensity at both angles and stronger polarization at $10^\circ$, and turbulence exhibits filamentary emission \citep{upreti2024}, obscuring the twisted structure. In the bottom recollimation region, several bright knotty structures are seen at $10^\circ$, which correspond to the recollimation shocks. Flows on the jet surface move more orderly, although the bright knots are also obscured. Therefore, it is seen that the EVPA patterns are perpendicular to the jet propagation direction at $90^\circ$. This indicates a poloidal magnetic field at the edge of the jet. 

Notably, the broken power-law distribution shows a different profile at $10^\circ$. 
We can even see the inner bright knots. The EVPA patterns are parallel with the jet propagation direction at the jet spine, and perpendicular to it at the edges. These features are consistent with the inner jet emission at different frequencies (see Figures~\ref{fig:freq}(b,c), which show the inner structure of the jet). This is attributed to the low fraction of medium and high-energy electrons in the broken power-law eDF, which results in a low optical depth. A limb-brightening morphology is observed at $90^\circ$ for three eDFs. The largest brightness ratios across the transverse jet (i.e., the brightness ratios between limb and jet core) are 1.4, 1.8, and 2.6 for the thermal, $\kappa$, and power-law eDFs. Notice that the brightness ratios for non-thermal distributions are higher, indicating they enhance the emission of the ambient medium. 

To investigate the effect of ambient medium on emission signatures, we compare emission at 86 GHz in Figures~\ref{fig:angle}(e) and Figure~\ref{fig:freq}(d) for viewing angle $10^\circ$ and $\kappa$ distribution. The emission shown in Figure 3(d) and also in appendix~\ref{90} does not include the ambient medium, identified via a tracer. The absence of the ambient medium in modifying the emission signature is very much visible. The turbulent features and limb brightened structures observed in Figure~\ref{fig:angle}(e) are absent for Figure~\ref{fig:freq}(d). Instead, we see the presence of inner knots and smooth outer edges. For the rest of this work, we include the emission from the ambient environment (no elimination by jet tracer).

\subsection{Influence of frequencies}

Due to frequency-dependent self-absorption, the optical depth of the plasma varies with the observing frequency, thereby altering the observed structure of the jet. To explore how frequencies affect radiation structure, we compare the ray-traced images at $10^\circ$ for 1.5~GHz, 160~$\mu$m, 521~nm, 1.2~keV, with those at 86~GHz (Figure~\ref{fig:freq}). The ray-traced images at $90^\circ$ are also calculated (see Appendix~\ref{90}).
Considering the similar general conclusions obtained from different eDFs, here we focus our discussion on the results from the $\kappa$ distribution.

At 1.5~GHz, as is shown in Figure~\ref{fig:freq}(a), the jet is completely obscured by optically thick expanding ambient medium, and only the emission from the jet head is seen from a small viewing angle.
On the contrary, at 521~nm and 1.2~keV (Figure~\ref{fig:freq}(b,c)), both the bright knots caused by recollimation shocks and the twist kink are clearly shown. These structures can also be retrieved at 86~GHz with a tracer to exclude the environment (Fig.~\ref{fig:freq}(d)). We note that the knot features in the jet become more diffuse at lower frequency.
 
Along the jet spine, the EVPA pattern is nearly parallel to the jet propagation direction, whereas at the edges of the jet, it becomes perpendicular. Our result agrees well with that found in \cite{kramer2021}. 
This polarization feature arises from distinct magnetic field configurations in the emission region between the central and jet edge. While the axial EVPA pattern originates from the jet inner component where the toroidal magnetic field component dominates, the poloidal magnetic field plays a predominant role in the emission from the jet boundary (Figure~\ref{fig:angle}(d-i)). Consequently, this magnetic field dichotomy naturally produces different EVPA patterns at the jet center and edge.

\section{Time Series Analysis} \label{tsa}

\begin{figure}[!h]
    \centering
    \includegraphics[width=0.5\textwidth]{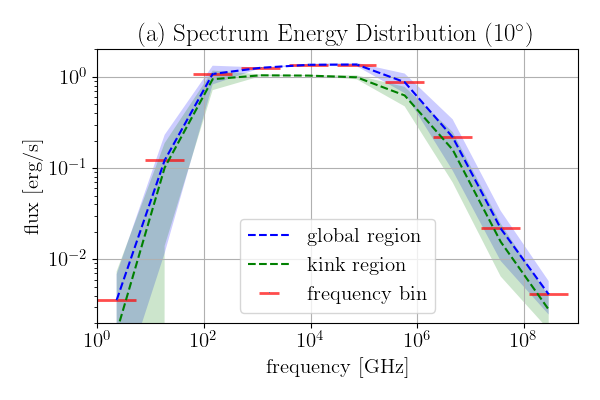}
    \caption{Time-average spectrum energy distribution at an inclination angle $i=10^\circ$. We make an average every 5 frequencies and mark the red horizontal lines. The whole jet region and the specified kink region are plotted in blue and green dashed lines, respectively. The light blue (green) shaded region indicates systematic uncertainty due to time variability and frequency bins.}
    \label{fig:sed}
\end{figure}

Next, we perform time series analyses.
We calculate multi-wavelength light curves between $t_s=300$ and $t_s=1100$. The output cadence is $t_s=2$. For each epoch, we compute the integrated polarized flux of 50 frequencies distributed evenly over a logarithmic domain from 1~GHz to $10^9$~GHz. The viewing angle is fixed to $10^\circ$ to apply to blazars.

\subsection{Spectrum Energy Distribution}

Figure~\ref{fig:sed} shows the spectral energy distribution of the over-pressured magnetized jet.
We set every 5 frequencies as a bin and make an average among these frequencies. Therefore, there are 10 characteristic frequencies in the spectral energy distribution (SED). 
We compute the SEDs for both the global region and the kink region ($40<z<60$), respectively. As illustrated in Figure~\ref{fig:tsa}(a), the flux rises rapidly from 1~GHz, reaching a peak range around $10^3$-$10^4$~GHz, followed by a sharp decline. 
The SED from the kink region follows the SED of the whole region, and the deviation is small.
It indicates that the multi-wavelength emission is dominated by the kink region.

\subsection{Seek QPO signatures from kink}

\begin{figure*}[ht]
	\centering
	\includegraphics[width=0.56\textwidth]{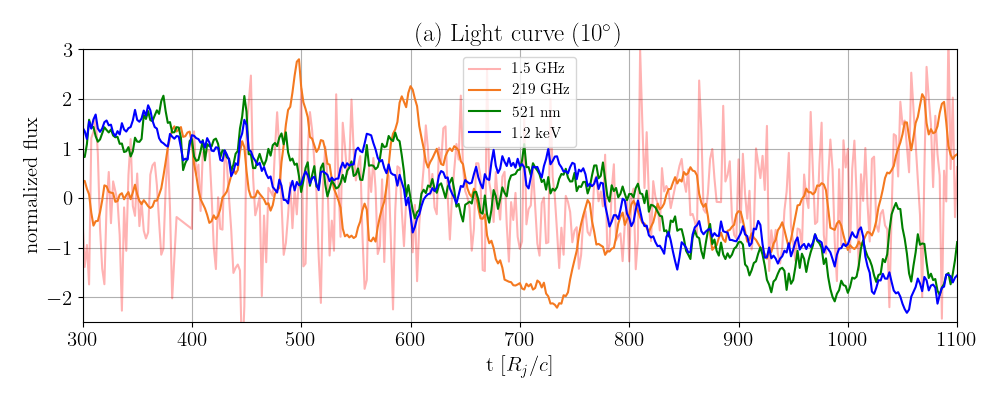}
	\includegraphics[width=0.34\textwidth]{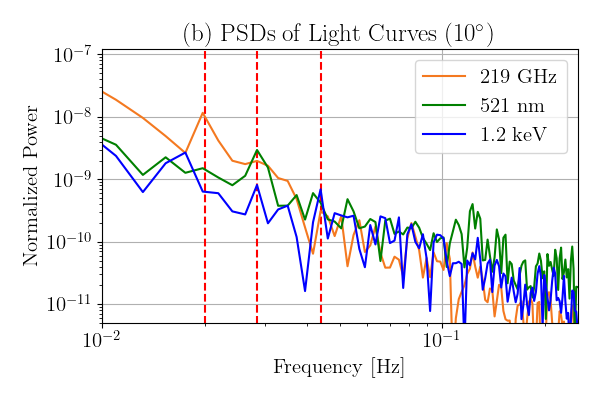}
	\includegraphics[width=0.56\textwidth]{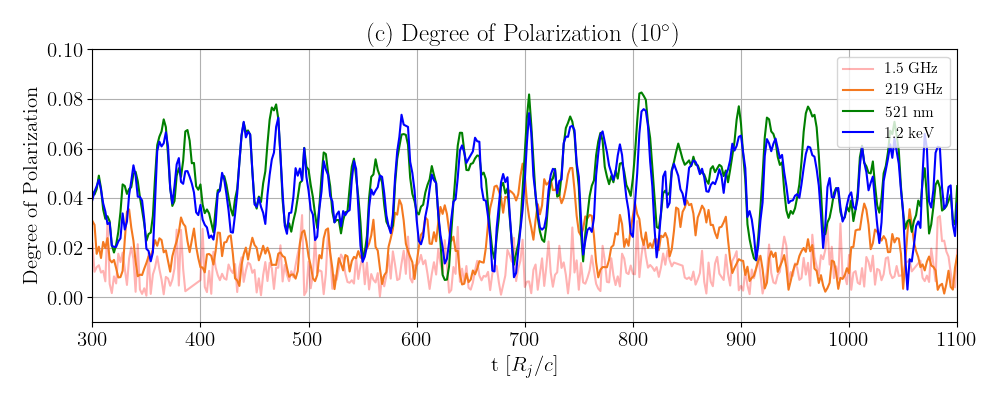}
    \includegraphics[width=0.34\textwidth]{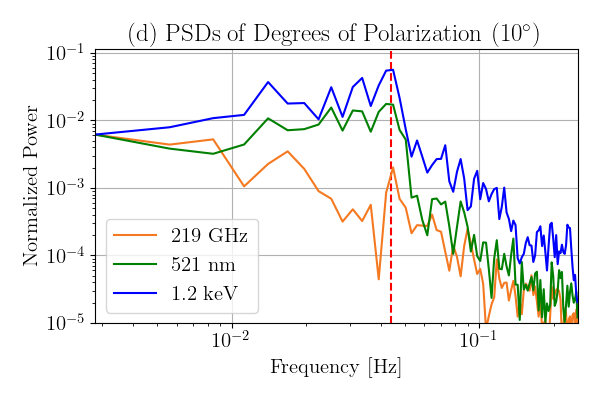}
    \includegraphics[width=0.35\textwidth]{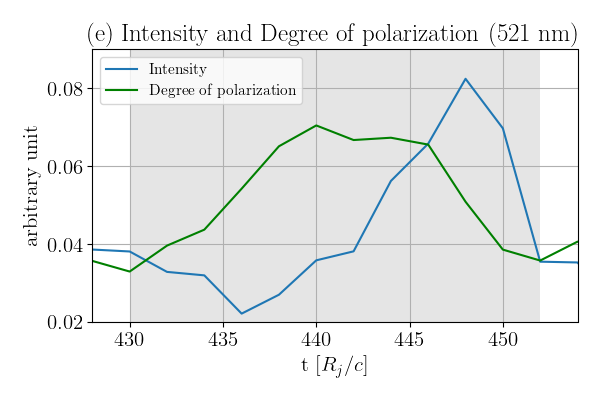}
    \includegraphics[width=0.26\textwidth]{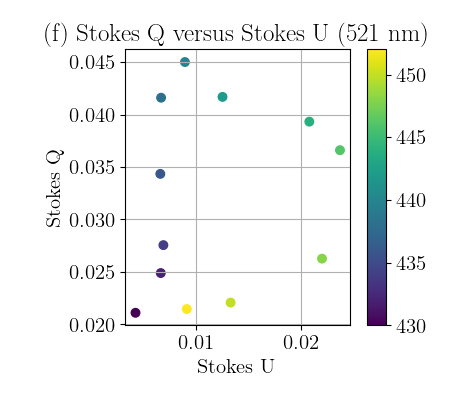}
	\caption{Panel (a) shows the light curves of normalized intensity at four representative frequencies: 1.5~GHz (pink), 219~GHz (orange), 521~nm (green), and 1.2~keV (blue). Panel (b) presents PSDs of light curves, where a Hanning window of 228 segments is applied to smooth the signals. The vertical dashed red lines indicate the peak frequencies, which are 20~mHz, 29~mHz, and 44~mHz, respectively. Panel (c) plots the time evolution of the total polarization degree of the kink region at four frequencies. Panel (d) shows the corresponding PSDs of degrees of polarization with a Hanning window of 178 segments. The vertical dashed red line marks 44~mHz. All of the PSDs are normalized at the lowest frequency. 
    Panel (e) shows the intensity and degree of polarization at 521~nm during a time interval with $t_s=428$-$456$. The grey shaded interval ($t_s=430$-$454$) is used for plotting panel (f). Panel (f) plots the Q-U plane for 521~nm. The colors of scattered points change over time. }
	\label{fig:tsa}
\end{figure*}

It is expected that the rotation of the kink structure can drive QPOs. 
Following \cite{dong2020}, we restrict the emission to the kink region and select four characteristic frequencies: 1.5~GHz, 219~GHz, 521~nm, and 1.2~keV. Figure~\ref{fig:tsa}(a) shows their light curves in normalized flux. 

The emissions from our chosen different frequencies represent different physical properties of jets. 
1.5~GHz emission traces the diffuse ambient medium around the jet, and it is gradually saturated at later simulation times. 
The emission at 219~GHz is dominated by the turbulence at the jet interface. It shows a large variation.
The images of 521~nm  and 1.2~keV are proxies of the inner structure of the jet. Due to the development of a kink structure by instability, QPOs are seen.

In Figure~\ref{fig:tsa}(b), we plot the PSDs of 219~GHz, 521~nm, and 1.2~keV. A peak at 44~mHz is only detected at 1.2~keV, and another peak at 29~mHz is detected at both 1.2~keV and 521~nm. We naturally connect them with the rotation periods of the kink structure, 27 and 46~mHz (see Appendix~\ref{kink}). However, at 219~GHz, which describes the emission from the medium surrounding the jet, has a unique peak at 20~mHz.
It is noted that the 219~GHz emission originates from turbulence surrounding the jet rather than the kink structure itself. Therefore, as the kink instability evolves, portions of the jet component transit to turbulence. This transition results in emission that continuously oscillates at a lower frequency relating to the peak at 20~mHz.
Notice that we set $M_{BH}=6.5\times10^7 M_\odot$, the three periods correspond to 8.5, 12.9, and 18.7~days.
We fit the PSDs over the 100-10~mHz interval with a power law, $P \propto f^{-\alpha}$, yielding $\alpha = 2.8$ at 219~GHz, $\alpha = 1.6$ at 521~nm, and $\alpha = 1.5$ at 1.2~keV. 
We construct a hypothesis test to identify the significance of the QPO peaks \citep{percival1993,vaughan2005} (see appendix~\ref{sign} for more detailed calculation). The P-values of the peaks are $\lesssim0.01$, except for the 29~mHz peak at 1.2~keV (P-value=0.11). It indicates that, except for the 29~mHz peak at 1.2~keV, QPO peaks are statistically significant.

\subsection{Degree of Polarization}

In Figure~\ref{fig:tsa}(c), we show the time evolution of degrees of linear polarization for four characteristic frequencies at the kink region of an over-pressured magnetized jet. The degree of polarization is calculated as total linear polarization emission divided by total intensity.
The degrees are positively correlated to frequency. 
At 1.5~GHz, the degree of polarization is very low, consistent with Figure~\ref{fig:freq}(a).
At 219~GHz, the degree varies around 0.02.
At 521~nm and 1.2~keV, the degree oscillates between 0.02 and 0.08.

We also show the corresponding PSDs in Figure~\ref{fig:tsa}(d). The degree of polarization exhibits regular oscillatory behavior. All frequencies show a peak range around 40-50 mHz, which confirms the periodic variation of degrees of polarization and is consistent with the rotation period of the kink structure (46~mHz).

We investigate the correlation of time variability between intensity and degree of polarization. 
Here we calculate Pearson's correlation coefficient, which measures linear correlation between two sets of data. The coefficients are -0.36 ($\text{P-value}=1.2\times10^{-13}$), 0.026 ($\text{P-value}=0.60$), -0.09 ($\text{P-value}=6.2\times10^{-2}$) in the low-frequency to high-frequency range. The least coefficient is below -0.3, indicating that dissipation of the toroidal magnetic field as the kink develops \citep{zhang2017}, which confirms the anti-correlation in \citep{dong2020}.

During the rotation of the kink structure, the polarization angle exhibits a synchronous swing. We show a Q-U plot (Figure~\ref{fig:tsa}f) for a period $t_s=430$-$454$ (Figure~\ref{fig:tsa}e) at 521 nm. The trajectory traces a clockwise rotation. This indicates that while the initial emission is dominated by the toroidal magnetic field, the contribution from the poloidal magnetic component is enhanced during the period of strong intensity \citep{dong2020}.

Lastly, we analyze time domain signals at a viewing angle of $90^\circ$. Degrees of linear polarization are generally higher at $90^\circ$. QPOs at the same frequencies are detected in intensity and linear polarization. However, we derive a positive correlation instead. The positive correlation coefficient gradually weakens with increasing frequencies, suggesting that it depends on frequency and viewing angle (see Appendix~\ref{90}).

\section{Discussion and Conclusion} \label{con}

In this work, we newly develop the SRRT code {\tt RaptorP} based on the polarized GRRT code {\tt RAPTOR}, which includes polarized synchrotron emission, absorption, and Faraday rotation under Minkowski metric. For eDFs, {\tt RaptorP} comprehensively integrates PIC simulation results, incorporating the effects of magnetic reconnection and turbulence to help a more informed estimation of the non-thermal electron population. 

From the polarized images of an over-pressured magnetized jet, we confirm the presence of a toroidal magnetic field in the inner jet, while the emission from the surrounding medium shows a dominance of poloidal magnetic fields. The features of the kink and recollimation shocks in the jet can be detected from intermediate viewing angles. The limb-brightening morphology is demonstrated at large viewing angles, which arises from jet-ambient medium interactions.

Through the comparisons of multi-wavelength polarization images, we reveal the dependence of the radiation profile on frequency. 
At a lower radio frequency of 1.5~GHz, only the expanding ambient gas is observed. 
At a higher radio frequency of 86~GHz, turbulence around the jet is observed. The kink region dominates the emission. The EVPA patterns are perpendicular to the propagation direction.
At 521~nm and 1.2 keV, the inner structure of the jet is seen, including the bright knots and twisted kinks. The EVPA patterns at the jet axis are parallel with the jet propagation direction, and at the jet edge, become perpendicular to it. These results confirm a stratified jet \citep{perlman1999,gesu2023,dawoon2024,kim2024}.

We derive the SED of the jet, and analyze QPOs of total intensity and linear polarization degree at 219~GHz, 521~nm, and 1.2~keV from both $10^\circ$ and $90^\circ$. The QPO frequencies differ at different frequencies. At higher frequencies, such as 521~nm and 1.2~keV, the intrinsic rotation period of the kink structure in the jet is detected, while the lower frequency (219~GHz) shows a low-frequency oscillation, which comes from the surrounding turbulent medium. Although our observed inclination angle is still larger, we confirmed the QPO feature is naturally observed from the kink developed by instability after passing recollimation shocks in a magnetized jet that is seen in observations of the BL Lac jet \citep{jortstad2022}. In particular, we predict a characteristic clockwise rotation loop in the Q-U plane, associated with QPOs in blazars.

Besides, we also find that the correlation between the intensity and degrees of linear polarization depends on frequency and viewing angle. At a mostly pole-on view ($10^\circ$), the anti-correlation is confirmed. However, at edge-on view ($90^\circ$), we see a positive correlation coefficient that gradually weakens with increasing frequency. This effect is related to optical depth. We expect the correlation to reverse as the frequency increases.

An important caveat is that the multi-wavelength images depend sensitively on the $M_{unit}$ and $M_{BH}$. Moreover, the optical depth would be different for other setups of simulations. However, the qualitative conclusion will be invariant.

Another drawback of our SRRT calculation is the lack of time-lag, i.e., slow-light effect. Although \cite{bronzwaer2018} compared the fast-light approximation with the slow-light one and stated that the difference should be less than $5\%$. The scale of a jet is further larger than that of a black hole. Thus, we expect that the time-lag will affect the radiation image and the light curve, as pointed out by \cite{dong2020}. Recently \cite{Saiz-Perez2025} has shown the difference between fast-light and slow-light approximation on jet image based on SRHD jet propagation simulations. They show that the rotation and broadening of the shocks in relativistic jets are caused by the light travel time delay between the near side (front) and the far side (back) by the slow-light effect.
We will add the slow-light effect to the code in future work. Except for synchrotron radiation, inverse Compton Scattering contributes another peak from X-ray to $\gamma$-ray bands in jet SEDs \citep{blandford2019}. In this work, we do not consider it. Thus, the X-ray emission result is under-estimated. We will involve this effect in future work. We believe {\tt RaptorP} will be a powerful tool for exploring the properties of the relativistic jet, building a bridge between numerical simulations and observations. 

\begin{acknowledgements}
This research is supported by the National Key R\&D Program of China (Grant No.\,2023YFE0101200), the National Natural Science Foundation of China (Grant No.\,12273022, 12192220, 12133008), and the Shanghai Municipality Orientation Program of Basic Research for International Scientists (Grant No.\,22JC1410600). The simulations were performed on TDLI-Astro, Pi2.0, and Siyuan Mark-I at Shanghai Jiao Tong University. 
\end{acknowledgements}


\begin{thebibliography}{}
	
	\bibitem[Blandford et al.(2019)]{blandford2019}
	Blandford, R., Meier, D., \& Readhead, A.\ 2019, \araa, 57, 467
	
	\bibitem[Jiang et al.(2025)]{2025arXiv250712789J}
	Jiang, H.-X., Mizuno, Y., Dihingia, I.~K., et al.\ 2025, arXiv e-prints, arXiv:2507.12789
	
	\bibitem[Yang et al.(2024)]{Yang2024}
	Yang, H., Yuan, F., Li, H., et al.\ 2024, Science Advances, 10, adn3544
	
	\bibitem[Cruz-Osorio et al.(2022)]{Cruz-Osorio2022a}
	Cruz-Osorio, A., Fromm, C.~M., Mizuno, Y., et al.\ 2022, Nature Astronomy, 6, 103
	
	\bibitem[Blandford \& Znajek(1977)]{blandfords1977}
	Blandford, R.~D., \& Znajek, R.~L.\ 1977, \mnras, 179, 433
	
	\bibitem[Pushkarev et al.(2009)]{pushkarev2009}
	Pushkarev, A.~B., Kovalev, Y.~Y., Lister, M.~L., \& Savolainen, T.\ 2009, \aap, 507, L33
	
	\bibitem[Oei et al.(2024)]{oei2024}
	Oei, M.~S.~S.~L., Hardcastle, M.~J., Timmerman, R., et al.\ 2024, \nat, 633, 537
	
	\bibitem[Porth et al.(2017)]{porth2017}
	Porth, O., Olivares, H., Mizuno, Y., et al.\ 2017, Computational Astrophysics and Cosmology, 4, 1
	
	\bibitem[Prather(2024)]{prather2024}
	Prather, B.~S.\ 2024, arXiv e-prints, arXiv:2408.01361
	
	\bibitem[Mignone et al.(2007)]{mignone2007}
	Mignone, A., Bodo, G., Massaglia, S., et al.\ 2007, \apjs, 170, 228
	
	\bibitem[Ayache et al.(2022)]{ayache2022}
	Ayache, E.~H., van Eerten, H.~J., \& Eardley, R.~W.\ 2022, \mnras, 510, 1315
	
	\bibitem[Tchekhovskoy \& Bromberg(2016)]{tchekhovskoy2016}
	Tchekhovskoy, A., \& Bromberg, O.\ 2016, \mnras, 461, L46
	
	\bibitem[Savard et al.(2025)]{savard2025}
	Savard, K., Matthews, J.~H., Fender, R., \& Heywood, I.\ 2025, \mnras, 540, 1084
	
	\bibitem[Mizuno et al.(2015)]{mizuno2015}
	Mizuno, Y., Gómez, J.~L., Nishikawa, K.-I., et al.\ 2015, \apj, 809, 38
	
	\bibitem[Thimmappa et al.(2024)]{thimmappa2024}
	Thimmappa, R., Neilsen, J., Haggard, D., et al.\ 2024, \apj, 969, 128
	
	\bibitem[Mizuno et al.(2014)]{mizuno2014}
	Mizuno, Y., Hardee, P.~E., \& Nishikawa, K.-I.\ 2014, \apj, 784, 167
	
	\bibitem[Mizuno et al.(2009)]{mizuno2009}
	Mizuno, Y., Lyubarsky, Y., Nishikawa, K.-I., \& Hardee, P.~E.\ 2009, \apj, 700, 684
	
	\bibitem[Millas et al.(2017)]{millas2017}
	Millas, D., Keppens, R., \& Meliani, Z.\ 2017, \mnras, 470, 592
	
	\bibitem[Matsumoto \& Masada(2019)]{matsumoto2019}
	Matsumoto, J., \& Masada, Y.\ 2019, \mnras, 490, 4271
	
	\bibitem[Gourgouliatos \& Komissarov(2018)]{gourgouliatos2018}
	Gourgouliatos, K.~N., \& Komissarov, S.~S.\ 2018, Nature Astronomy, 2, 167
	
	\bibitem[Abolmasov \& Bromberg(2023)]{abolmasov2023}
	Abolmasov, P., \& Bromberg, O.\ 2023, \mnras, 520, 3009
	
	\bibitem[Toma et al.(2017)]{toma2017}
	Toma, K., Komissarov, S.~S., \& Porth, O.\ 2017, \mnras, 472, 1253
	
	\bibitem[Hu et al.(2025)]{hu2025}
	Hu, X.-F., Mizuno, Y., \& Fromm, C.~M.\ 2025, \aap, 693, A154
	
	\bibitem[Bronzwaer et al.(2018)]{bronzwaer2018}
	Bronzwaer, T., Davelaar, J., Younsi, Z., et al.\ 2018, \aap, 613, A2
	
	\bibitem[Bronzwaer et al.(2020)]{bronzwaer2020}
	Bronzwaer, T., Younsi, Z., Davelaar, J., \& Falcke, H.\ 2020, \aap, 641, A126
	
	\bibitem[Younsi et al.(2012)]{younsi2012}
	Younsi, Z., Wu, K., \& Fuerst, S.~V.\ 2012, \aap, 545, A13
	
	\bibitem[Event Horizon Telescope Collaboration et al.(2024)]{eht2024}
	Event Horizon Telescope Collaboration, Akiyama, K., Alberdi, A., et al.\ 2024, \apjl, 964, L26
	
	\bibitem[Dubey et al.(2023)]{dubey2023}
	Dubey, R.~P., Fendt, C., \& Vaidya, B.\ 2023, \apj, 952, 1
	
	\bibitem[Dubey et al.(2024)]{dubey2024}
	Dubey, R.~P., Fendt, C., \& Vaidya, B.\ 2024, \apj, 976, 144
	
	\bibitem[Kramer \& MacDonald(2021)]{kramer2021}
	Kramer, J.~A., \& MacDonald, N.~R.\ 2021, \aap, 656, A143
	
	\bibitem[Kramer et al.(2024)]{kramer2024}
	Kramer, J.~A., MacDonald, N.~R., Paraschos, G.~F., \& Ricci, L.\ 2024, \aap, 691, A14
	
	\bibitem[Vaidya et al.(2018)]{vaidya2018}
	Vaidya, B., Mignone, A., Bodo, G., et al.\ 2018, \apj, 865, 144
	
	\bibitem[Dong et al.(2020)]{dong2020}
	Dong, L., Zhang, H., \& Giannios, D.\ 2020, \mnras, 494, 1817
	
	\bibitem[Jorstad et al.(2022)]{jortstad2022}
	Jorstad, S.~G., Marscher, A.~P., Raiteri, C.~M., et al.\ 2022, \nat, 609, 265
	
	\bibitem[Dullemond et al.(2012)]{dullemond2012}
	Dullemond, C.~P., Juhasz, A., Pohl, A., et al.\ 2012, Astrophysics Source Code Library, ascl:1202.015
	
	\bibitem[Alves et al.(2018)]{alves2018}
	Alves, E.~P., Zrake, J., \& Fiuza, F.\ 2018, Physical Review Letters, 121, 245101
	
	\bibitem[Ball et al.(2018)]{ball2018}
	Ball, D., Sironi, L., \& Özel, F.\ 2018, \apj, 862, 80
	
	\bibitem[Meringolo et al.(2023)]{meringolo2023}
	Meringolo, C., Cruz-Osorio, A., Rezzolla, L., \& Servidio, S.\ 2023, \apj, 944, 122
	
	\bibitem[Chatterjee et al.(2021)]{chatterjee2021}
	Chatterjee, K., Markoff, S., Neilsen, J., et al.\ 2021, \mnras, 507, 5281
	
	\bibitem[Gammie \& Popham(1998)]{gammie1998}
	Gammie, C.~F., \& Popham, R.\ 1998, \apj, 498, 313
	
	\bibitem[Davelaar et al.(2019)]{davelaar2019}
	Davelaar, J., Olivares, H., Porth, O., et al.\ 2019, \aap, 632, A2
	
	\bibitem[Xiao(2006)]{xiao2006}
	Xiao, F.\ 2006, Plasma Physics and Controlled Fusion, 48, 203
	
	\bibitem[Fromm et al.(2022)]{fromm2022}
	Fromm, C.~M., Cruz-Osorio, A., Mizuno, Y., et al.\ 2022, \aap, 660, A107
	
	\bibitem[Saiz-Pérez et al.(2025)]{Saiz-Perez2025}
	Saiz-Pérez, A., Fromm, C.~M., Perucho, M., et al.\ 2025, \aap, 693, A169
	
	\bibitem[Upreti et al.(2024)]{upreti2024}
	Upreti, N., Vaidya, B., \& Shukla, A.\ 2024, Journal of High Energy Astrophysics, 44, 146
	
	\bibitem[Sciaccaluga et al.(2025)]{sciaccaluga2025}
	Sciaccaluga, A., Costa, A., Tavecchio, F., et al.\ 2025, \aap, 699, A296
	
	\bibitem[Böttcher \& Dermer(2010)]{bottcher2010}
	Böttcher, M., \& Dermer, C.~D.\ 2010, \apj, 711, 445
	
	\bibitem[Sironi et al.(2015)]{sironi2015}
	Sironi, L., Petropoulou, M., \& Giannios, D.\ 2015, \mnras, 450, 183
	
	\bibitem[Davelaar et al.(2020)]{davelaar2020}
	Davelaar, J., Philippov, A.~A., Bromberg, O., \& Singh, C.~B.\ 2020, \apjl, 896, L31
	
	\bibitem[Tripathi et al.(2021)]{tripathi2021}
	Tripathi, A., Gupta, A.~C., Aller, M.~F., et al.\ 2021, \mnras, 501, 5997
	
	\bibitem[Sharma et al.(2025)]{sharma2025}
	Sharma, A., Chaudhary, S., Sarath, A., \& Bose, D.\ 2025, arXiv e-prints, arXiv:2505.23697
	
	\bibitem[Kang et al.(2015)]{kang2015}
	Kang, S., Lee, S.-S., \& Byun, D.-Y.\ 2015, Journal of Korean Astronomical Society, 48, 257
	
	\bibitem[Perlman et al.(1999)]{perlman1999}
	Perlman, E.~S., Biretta, J.~A., Zhou, F., Sparks, W.~B., \& Macchetto, F.~D.\ 1999, \aj, 117, 2185
	
	\bibitem[Liodakis et al.(2022)]{liodakis2022}
	Liodakis, I., Marscher, A.~P., Agudo, I., et al.\ 2022, Nature, 611, 677
	
	\bibitem[Di Gesu et al.(2023)]{gesu2023}
	Di Gesu, L., Marshall, H.~L., Ehlert, S.~R., et al.\ 2023, Nature Astronomy, 7, 1245
	
	\bibitem[Kim et al.(2024)]{dawoon2024}
	Kim, D.~E., Di Gesu, L., Liodakis, I., et al.\ 2024, \aap, 681, A12
	
	\bibitem[Kim et al.(2024)]{kim2024}
	Kim, D.~E., Di Gesu, L., Marin, F., et al.\ 2024, Galaxies, 12, 20
	
	\bibitem[Vaughan(2005)]{vaughan2005}
	Vaughan, S.\ 2005, \aap, 431, 391
	
	\bibitem[Percival \& Walden(1993)]{percival1993}
	Percival, D.~B., \& Walden, A.~T.\ 1993, Spectral Analysis for Physical Applications
	
	\bibitem[Chatterjee et al.(2019)]{chatterjee2019}
	Chatterjee, K., Liska, M., Tchekhovskoy, A., \& Markoff, S.~B.\ 2019, \mnras, 490, 2200
	
	\bibitem[Lalakos et al.(2024)]{lalakos2024}
	Lalakos, A., Tchekhovskoy, A., Bromberg, O., et al.\ 2024, \apj, 964, 79
	
	\bibitem[Barniol Duran et al.(2017)]{duran2017}
	Barniol Duran, R., Tchekhovskoy, A., \& Giannios, D.\ 2017, \mnras, 469, 4957
	
	\bibitem[Zhang et al.(2017)]{zhang2017}
	Zhang, H., Li, H., Guo, F., \& Taylor, G.\ 2017, \apj, 835, 125
	
	\bibitem[Bodo et al.(2022)]{bodo2022}
	Bodo, G., Mamatsashvili, G., Rossi, P., \& Mignone, A.\ 2022, \mnras, 510, 2391
	
	\bibitem[Musso et al.(2024)]{musso2024}
	Musso, M., Bodo, G., Mamatsashvili, G., Rossi, P., \& Mignone, A.\ 2024, \mnras, 532, 4810
	
\end{thebibliography}

\appendix
\section{Polarized radiative transfer and ray-tracing in GRRT code {\tt RAPTOR}}\label{rt}

{\tt RAPTOR} solves GRRT equations in two main steps. The first step is to determine the null geodesic:
\begin{equation}
	k^\alpha\nabla_{\alpha}f^\mu=0\label{6}
\end{equation}
where $f^\mu$ is the polarization four-vector.
Taking equation (\ref{6}) and the basic properties of Stokes parameters, we have:
\begin{align}
 \left. \frac{\mathrm{d}}{\mathrm{d}\lambda}\right|_{S}f^{\mu}&=-\Gamma_{\alpha\rho}^{\mu}k^{\alpha}f^{\rho},\label{7} \\
 \left. {\frac{\mathrm{d}}{\mathrm{d}\lambda}}\right|_{S}I&=0,\label{8}\\
 \left. \frac{\mathrm{d}}{\mathrm{d}\lambda}\right|_{S}I_{\mathrm{pol}}&=0,\label{9}
\end{align}
where $\lambda$ is the affine parameter and the subscript $S$ implies that we only consider the propagation through the curved spacetime.

In the second step, the contribution of the plasma is included by performing radiative transfer calculations:
\begin{equation}\left. \frac{\mathrm{d}}{\mathrm{d}\lambda}\right|_P
	\begin{pmatrix}
		I \\
		Q \\
		\mathcal{U} \\
		\mathcal{V}
	\end{pmatrix}=
	\begin{pmatrix}
		j_I \\
		j_Q \\
		j_U \\
		j_V
	\end{pmatrix}-
	\begin{pmatrix}
		\alpha_I & \alpha_Q & \alpha_U & \alpha_V \\
		\alpha_Q & \alpha_I & \rho_V & -\rho_U \\
		\alpha_U & -\rho_V & \alpha_I & \rho_Q \\
		\alpha_V & \rho_U & -\rho_Q & \alpha_I
	\end{pmatrix}
	\begin{pmatrix}
		I \\
		Q \\
		\mathcal{U} \\
		\mathcal{V}
	\end{pmatrix},\end{equation}
where $j$, $\alpha$, and $\rho$ are the emission, absorption, and rotation coefficients, respectively. The subscript $P$ denotes the contribution from the ray's interaction with the plasma, ignoring the effects of spacetime propagation.

\section{Derivation of $w$ in the $\kappa$ distribution}\label{w}

\begin{figure}[!h]
	\centering
	\includegraphics[width=0.3\textwidth]{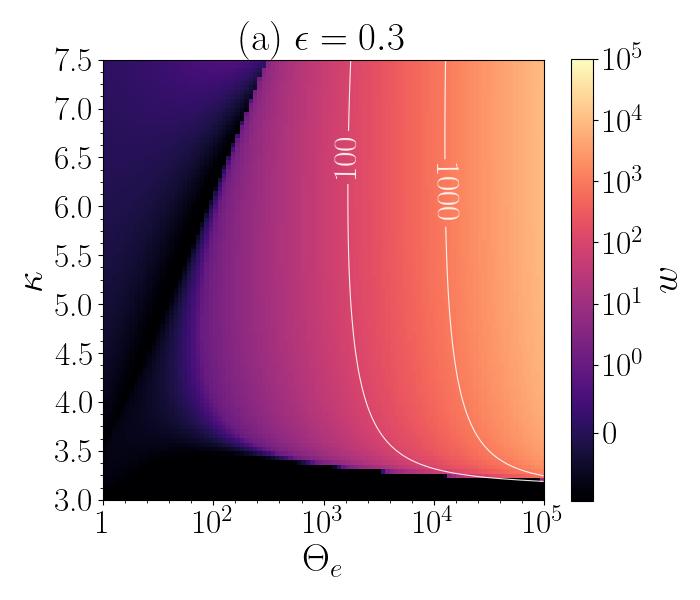}
	\includegraphics[width=0.3\textwidth]{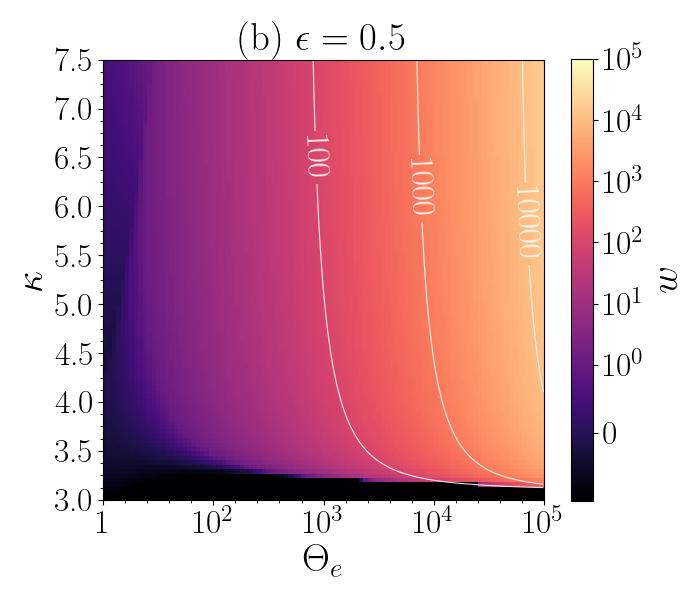}
	\includegraphics[width=0.3\textwidth]{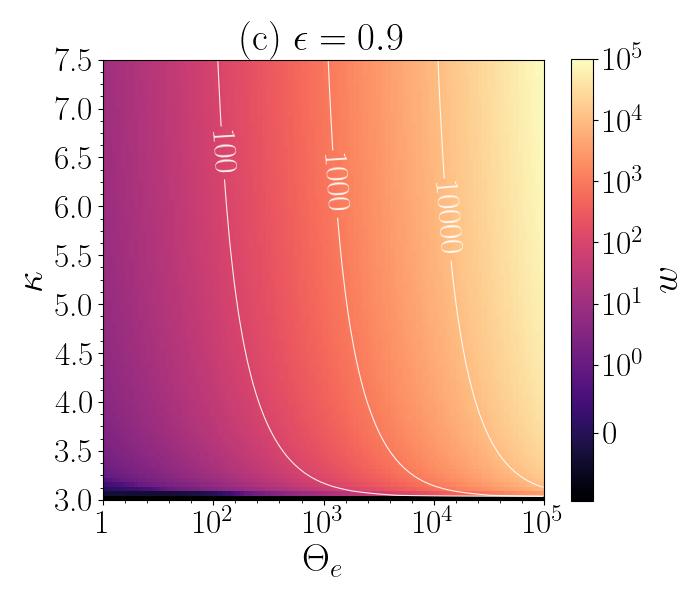}
	\caption{Dependency of $w$ on $\Theta_e$ and $\kappa$ at different efficiency $\epsilon=0.3$ ({\it a}), $0.5$ ({\it b}), and $0.9$ ({\it c}), which is the solution of Eq.~(\ref{b5}). Contour lines at $10^2$,$10^3$, and $10^4$ are plotted.}
	\label{fig:w}
\end{figure}

By employing $\gamma_0\gg1$, we rewrite Eq.~\ref{10} as
\begin{equation}
    \int_{\gamma_{0}}^{\infty}\frac{dN}{d\gamma}\gamma d\gamma=\frac{\int_{\gamma_{0}}^{\infty}f_{MJ}(\gamma,\Theta_e)\gamma d\gamma}{1-\epsilon}.\label{b1}
\end{equation}
and we simplify the expressions of the thermal distribution as $\frac{\gamma^2}{\Theta_eK_2(1/\Theta_e)}e^{-\gamma/\Theta_e}$ and $\kappa$ distribution as $\gamma^2(1+\frac{\gamma}{\kappa w})^{-(\kappa+1)}$. Then we rewrite Eq.~\ref{b1} with normalization:

\begin{align}
    \frac{\int_{\gamma_{0}}^{\infty}\gamma^3(1+\frac{\gamma}{\kappa w})^{-(\kappa+1)}}{\int_{\gamma_{0}}^{\infty}\gamma^2(1+\frac{\gamma}{\kappa w})^{-(\kappa+1)}}&=
    \frac{1}{1-\epsilon}\frac{\int_{\gamma_{0}}^{\infty}\frac{\gamma^3}{\Theta_eK_2(1/\Theta_e)}e^{-\gamma/\Theta_e} d\gamma}{\int_{\gamma_{0}}^{\infty}\frac{\gamma^2}{\Theta_eK_2(1/\Theta_e)}e^{-\gamma/\Theta_e} d\gamma}\\
    \frac{6 \gamma_{0} (w \kappa)^{2}+\gamma_{0}^3 (-2 + \kappa) (-1 + \kappa) + 3 w \gamma_{0}^2 (-1 + \kappa) \kappa +6 w^3 \kappa^2}
{(-3 + \kappa) [\gamma_{0}^2 (-1+\kappa) + 2 w^2 \kappa +2\gamma_{0} w \kappa]}&=\frac{1}{1-\epsilon}(3\Theta_e+\frac{\gamma_{0}^3}{\gamma_{0}^2+2\gamma_{0}\Theta_e+2\Theta_e^2})\label{b2}
\end{align}

Define $C$ equals the right hand of Eq.~\ref{b2}, and we reorganize Eq.~\ref{b2} into a cubic equation:

\begin{equation}
    a w^3+b w^2+c w+d=0,\label{b5}
\end{equation}
\begin{equation}
     \begin{cases}
a &= 6\kappa^2,\\
b &= 6\gamma_{0}\kappa^2-2\kappa C(-3+\kappa),\\
c &= 3\gamma_{0}^2 (-1 + \kappa)\kappa-(-3+\kappa)C\gamma_{0}\kappa,\\
d &= \gamma_{0}^3 (-2 + \kappa) (-1 + \kappa)-C(-3 + \kappa)\gamma_{0}^2 (-1+\kappa),
    \end{cases}
\end{equation}

To guarantee the existence of at least one positive real root, the condition $d<0$ is a sufficient condition:

\begin{align}
\gamma_{0}^3 (-2 + \kappa) (-1 + \kappa)&<C(-3 + \kappa)\gamma_{0}^2 (-1+\kappa)\\
\gamma_{0}(\kappa-2)&<C(\kappa-3)\\
\gamma_{0}\frac{\kappa-2}{\kappa-3}&<\frac{1}{1-\epsilon}(3\Theta_e+\frac{\gamma_{0}^3}{\gamma_{0}^2+2\gamma_{0}\Theta_e+2\Theta_e^2})\\
3\Theta_e\frac{\kappa-2}{\kappa-3}&<\frac{3\Theta_e}{1-\epsilon}(1+\frac{9}{17})\\
1+\frac{1}{\kappa-3}&<\frac{1.529}{1-\epsilon}\\
\kappa&>2+\frac{1.529}{0.529+\epsilon}\label{b4}
\end{align}

where $\gamma_{0}\approx3\Theta_e$ \citep{chatterjee2021} is applied. This equation implies that $\kappa>3$ at $\epsilon=1$ and $\kappa>4.89$ at $\epsilon=0$. In SRRT calculations, the condition is satisfied across over $99.5\%$ of the simulation domain, and we also set the floor of $w$. If Eq.~\ref{b4} has more than one positive root, we select the largest one.

We show the dependency of $w$ on $\Theta_e$ and $\kappa$ at three different efficiencies $\epsilon=0.3$, $0.5$, and $0.9$ in Figure~\ref{fig:w}. 
It indicates that $w$ is sensitively proportional to $\Theta_e$ and $\epsilon$ but weakly proportional to $\kappa$. With $\epsilon$ increasing, the lower limit of $\kappa$ gradually approaches 3.

\section{Ray-traced images at $45^\circ$ for three eDFs}\label{45}

\begin{figure*}
	\centering
	\includegraphics[width=0.6\linewidth]{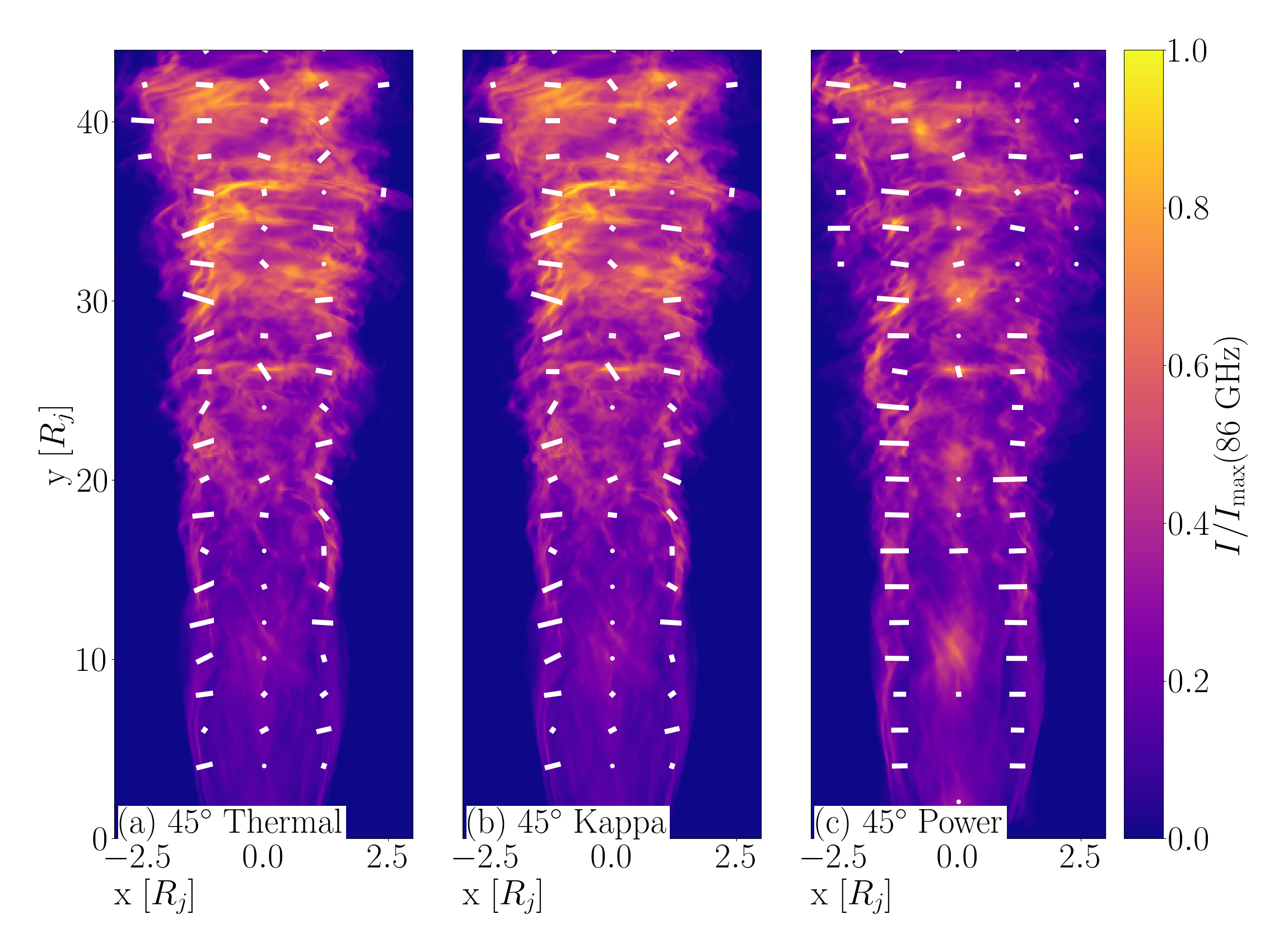}
	\caption{Same as Figure~\ref{fig:angle} but at an inclination angle of $45^\circ$.} 
	\label{fig:45}
\end{figure*}

We calculate the ray-traced images at a viewing angle of $45^\circ$ for thermal, $\kappa$, and broken power-law distributions (Figure~\ref{fig:45}), providing a direct comparison with \cite{kramer2021}. We do not retrieve the EVPA patterns observed in \cite{kramer2021} because the emission mostly comes from the ambient medium surrounding the jet. In addition, the inner bright knot is observed from the power-law eDF.

\section{The rotation period of the kink}\label{kink}

\begin{figure*}
	\centering
    \includegraphics[width=0.56\textwidth]{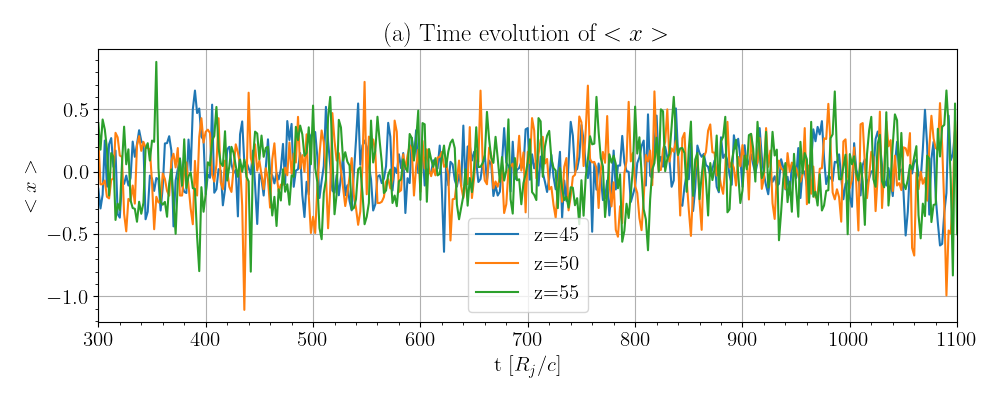}
	\includegraphics[width=0.34\textwidth]{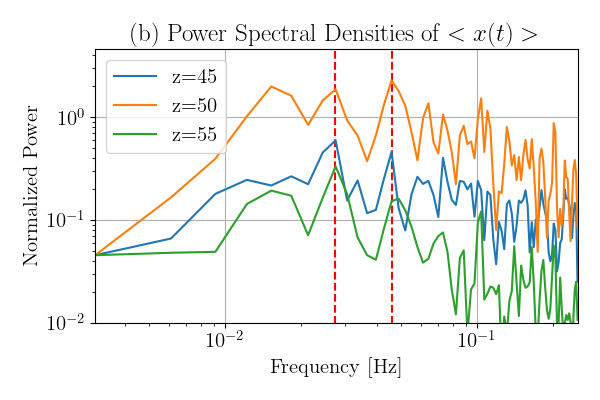}
    \caption{Panel (a) shows the average displacement of $x$  along the transverse cut of the jet ($y=0$) at $z= 40,45,50$. Panel (b) shows the corresponding PSDs of displacements with a Hanning window of 164 segments. The vertical dashed red lines mark the peak frequencies, which are 27 mHz and 46 mHz.}
	\label{fig:kink}
\end{figure*}

From 3D SRMHD simulations of over-pressured magnetized jets, we investigate the motion of the kink structure developed by CD kink instability.
To do it, We measure the average displacement of $x$ along the transverse cut of the jet ($y=0$) at $z=40$,$45$, and $50\,R_j$ (Figure~\ref{fig:kink}(a)) and the corresponding power spectral densities (PSDs) (Figure~\ref{fig:kink} (b)), where a Hanning window of 164 segments is applied to smooth the signals. From PSDs, two peaks at 27~mHz and 46~mHz are observed, indicating that the kink-like structure oscillates around the jet axis with certain periods.

\section{The significance test of QPO peaks}\label{sign}

To test the significance of QPO peaks, we fit the PSDs (QPO peak points are excluded) with a power-law formula, $P\propto f^{-\alpha}$, yielding $\alpha = 2.8$ at 219 GHz, $\alpha = 1.6$ at 521~nm, and $\alpha = 1.5$ at 1.2~keV, respectively. Then we construct a hypothesis test:
\begin{equation}
    \frac{P_{\text{obs}}}{P_{\text{model}}} \sim a\chi^2_\nu, \label{e1}
\end{equation}
where scale parameter $a$ and degree of freedom $\nu$ are unknown because we employ Hanning window smoothing. 
According to \cite{percival1993}, these two parameters can be estimated through
\begin{equation}
E(S) = a\nu, \quad \text{Var}(S) = 2a^2\nu. \label{e2}
\end{equation}

Since the mean and variance are derived from observation data, we obtain
\begin{equation}
\nu = \frac{2[E(S)]^2}{\text{Var}(S)}, \quad a = \frac{E(S)}{\nu}. \label{e3}
\end{equation}

Finally, we have $\frac{S(f)}{a} \sim \chi_\nu^2$ with P-value = $\Gamma\left(\frac{\nu}{2}, \frac{S}{2a}\right)$, where $\Gamma$ is the regularized upper incomplete gamma function. Using it, we calculate the P-values and significances of QPO peaks. As is illustrated in table~\ref{t1}, every peak is significant except the 29~mHz at 1.2~keV. 

\begin{table}
    \centering
    \begin{tabular}{lll}
    \hline \hline
        peak & P-value & $\sigma$ \\ \hline
        19 mHz at 219 GHz & $3\times10^{-3}$ & 2.75 \\ 
        29 mHz at 521 nm & $6.5\times10^{-7}$ & 4.84 \\
        29 mHz at 1.2 keV & 0.11 & 1.23 \\ 
        44 mHz at 1.2 keV & 0.011 & 2.29 \\ \hline
    \end{tabular}
    \caption{P-values and significances of the QPO peaks. The rows list peak frequencies. The first column shows the P-values. The second column shows the significances.}
    \label{t1}
\end{table}

\begin{figure*}
	\centering
	\includegraphics[width=0.6\linewidth]{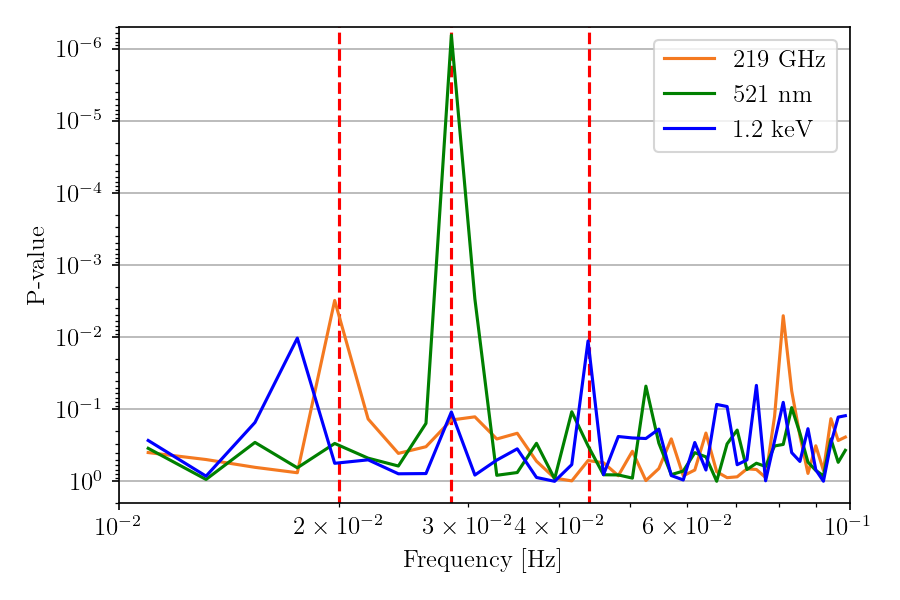}
	\caption{The P-values of the PSDs. The frequency range is from 10~mHz to 100~mHz. The red dashed lines mark 20~mHz, 29~mHz, 44~mHz.} 
	\label{fig:pvalue}
\end{figure*}
Furthermore, we plot the P-values of frequency points from 10~mHz to 100~mHz in Figure~\ref{fig:pvalue}.
It clearly shows distinguishable QPO peaks. Although we see the other peaks in the P-values plot (e.g., 17.5~mHz at 1.2~keV, 81~mHz at 219~GHz), these may be attributed to statistical factors: the Signal-Noise Ratio (SNR) at low frequency is worse, and the PSDs deviate from a simple power-law at high frequency. Therefore, we exclude these peaks.

\section{Effect on different inclination angle}\label{90}

\begin{figure*}
	\centering
	\includegraphics[width=0.85\linewidth]{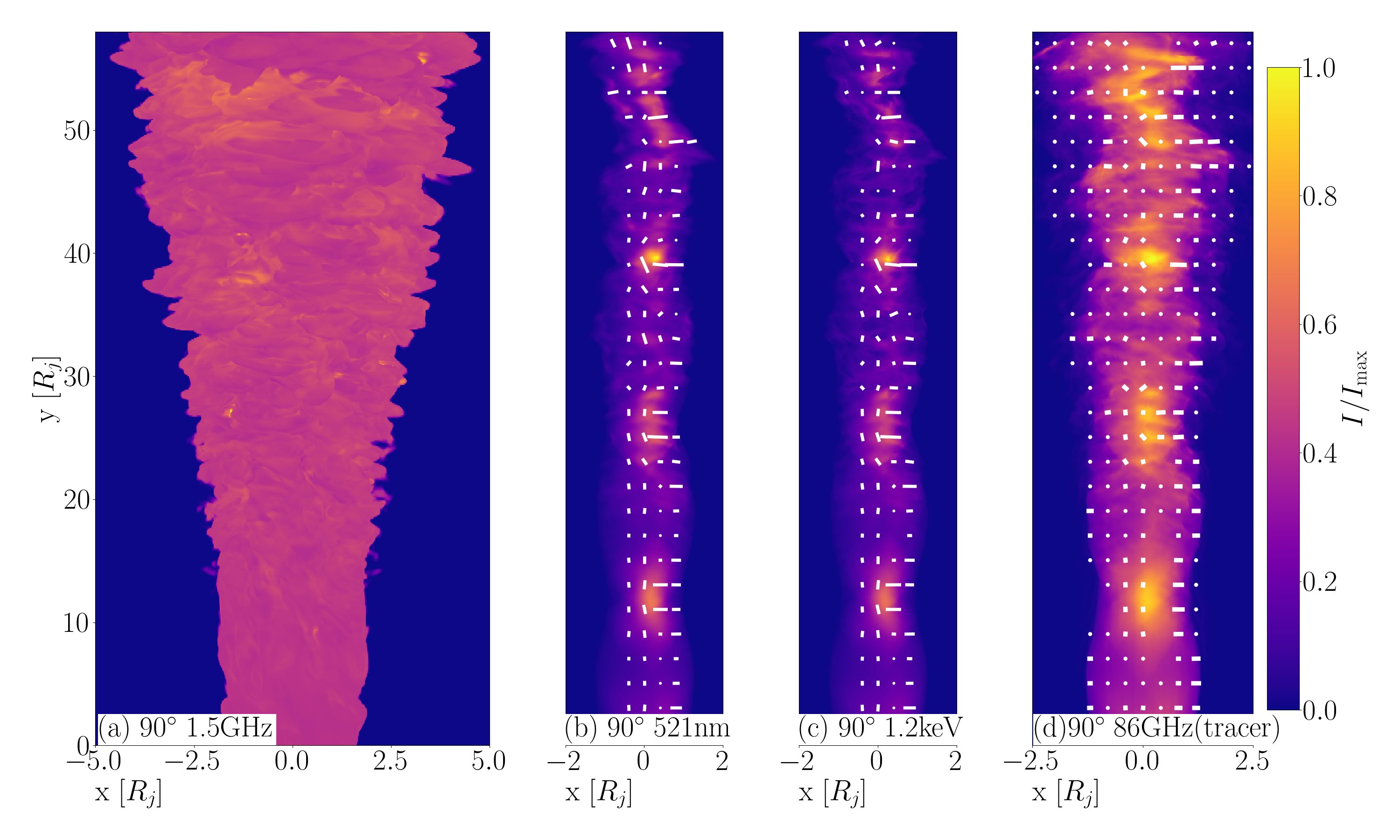}
	\caption{Same as Figure~\ref{fig:freq} but at a viewing angle of $90^\circ$.}
	\label{fig:freq90}
\end{figure*}

\begin{figure*}
    \centering
    \includegraphics[width=0.56\textwidth]{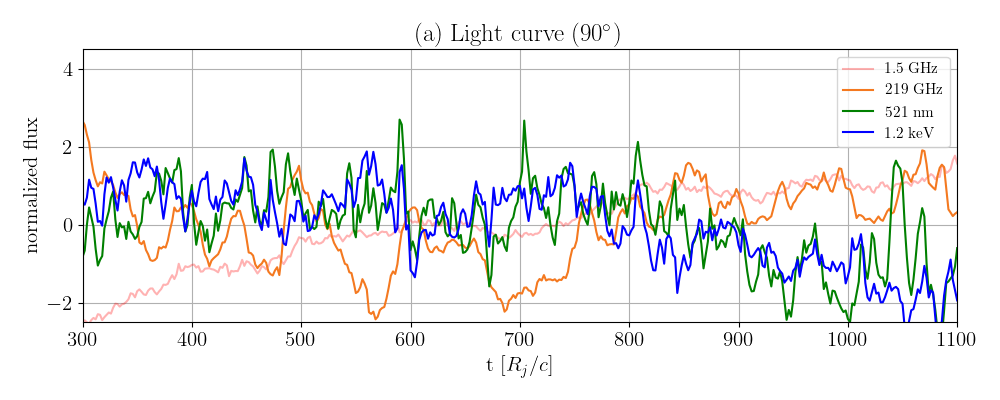}
	\includegraphics[width=0.34\textwidth]{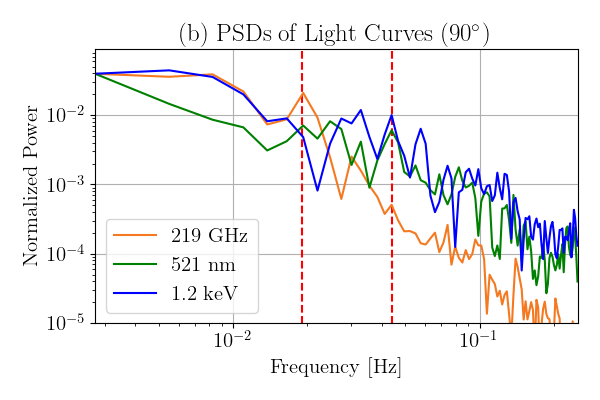}
	\includegraphics[width=0.56\textwidth]{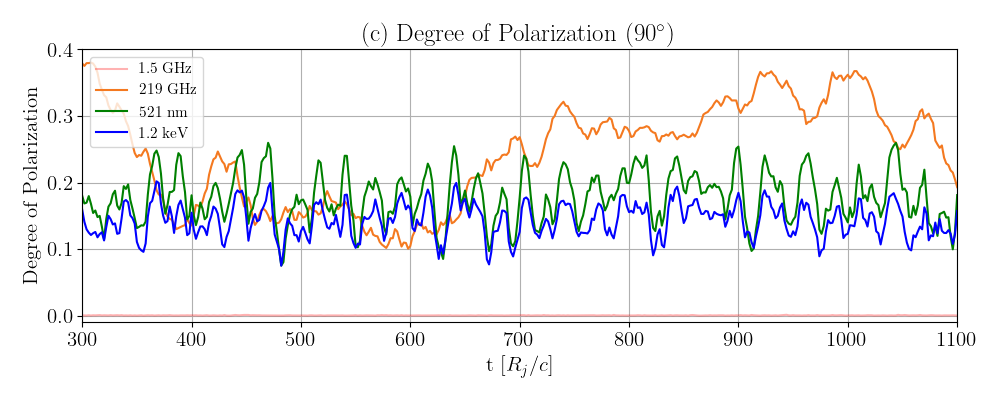}
    \includegraphics[width=0.34\textwidth]{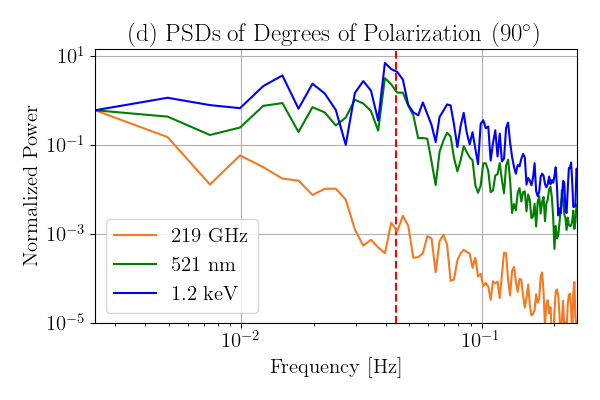}
    \caption{Same as Figure~\ref{fig:tsa} but at an inclination angle of $90^\circ$. In panel (b), a Hanning window of 182 segments are applied, and the vertical dashed red lines mark the peak frequencies, which are 19~mHz and 46~mHz. In panel (d),a Hanning window of 202 segments are applied, and the vertical dashed red line indicates 46~mHz.}
    \label{fig:tsa90}
\end{figure*}

To investigate the effect of the different inclination angles on time variability, we demonstrate the time series analysis of the edge-on view case ($90^\circ$) seen in Figure~\ref{fig:tsa90}. From this viewing angle, we also see the periodic oscillation. A peak at 46~mHz is detected at 1.2~keV and 521~nm, and the peak frequency at 19~mHz is invariant for 219~GHz (Figure~\ref{fig:tsa90}(b)). The degrees of linear polarization are generally higher at $90^\circ$ (Figure~\ref{fig:tsa90}(c)).
At 1.5~GHz, the degree of polarization is very low.
At 219~GHz, the degree varies dramatically between 0.1 and 0.4.
At 521~nm (1.2~keV), the degree oscillates between 0.15 (0.1) and 0.25 (0.2). Such oscillation behavior is similarly seen in \cite{dong2020}. 
The oscillated frequency (Figure~\ref{fig:tsa90}(d)) is the same as that in the $10^\circ$ case. It means the QPO is a robust feature among different inclination angles.

From the calculation of Pearson's correlation coefficients, we obtain 0.48 ($\text{P-value}=3.8\times10^{-24}$), 0.34 ($\text{P-value}=2.2\times10^{-12}$), 0.09 ($\text{P-value}=0.08$) from the low to high-frequency. It indicates a positive correlation of time variability between intensity and polarization degree at all frequencies.
These results seem to opposite sense reported by \cite{dong2020}. They suggested that the anti-correlation originates from the decline of $\sum u_t b_{\phi}^2/ \sum u_t b_{z}^2$ when the intensity increases. 
Thus, we additionally calculate the frequency-dependent correlation coefficients between $\sum u_t b_{\phi}^2/ \sum u_t b_{z}^2$ and intensity at each frequencies. In our case, we obtain 0.25 ($\text{P-value}=3.6\times10^{-7}$), 0.21 ($\text{P-value}=2.7\times10^{-5}$), -0.26 ($\text{P-value}=1.2\times10^{-7}$) from the low to high-frequency range. It means lower frequencies have a positive correlation.
We notice that \cite{dong2020} ignored the absorption of plasma. Thus, the frequency they adopted is effectively infinity. We can expect that the correlation between the total intensity and degree of polarization will have a turnover, as frequency increases.

\end{document}